\documentclass[12pt]{article}
\usepackage{color}
\usepackage{graphicx}
\def\nabstar#1{\nabla\kern-0.5pt\smash{\raise 4.5pt\hbox{$\ast$}}
               \kern-4.5pt_{#1}}

\def\drvstar#1{\partial\kern-0.5pt\smash{\raise 4.5pt\hbox{$\ast$}}
               \kern-5.0pt_{#1}}

\def\newline{\relax\ifhmode\null\hfil\break\else\nonhmodeerr@\newline\fi}
\def\frac#1#2{{#1\over#2}}
\def\text#1{{\hbox{\rm #1}}}
\def\flushpar{{\par \noindent}}

\newcommand{\beq}{\begin{equation}}
\newcommand{\eeq}{\end{equation}}
\newcommand{\bea}{\begin{eqnarray}}
\newcommand{\eea}{\end{eqnarray}}

\def\EQ{\hspace{-2mm} &=& \hspace{-2mm}}

\def\BA{\begin{eqnarray}}
\def\EA{\end{eqnarray}}
\def\BAN{\begin{eqnarray*}}
\def\EAN{\end{eqnarray*}}

\def\g5{\gamma_5}
\def\g4{\gamma_4}
\def\g3{\gamma_3}
\def\g2{\gamma_2}
\def\g1{\gamma_1}
\def\gi{\gamma_i}

\def\u{{\bf u}}

\def\s{{\bf s}}
\def\c{{\bf c}}
\def\q{{\bf q}}
\def\Q{{\bf Q}}
\def\ubar{\bar{\bf u}}

\def\sbar{\bar{\bf s}}
\def\cbar{\bar{\bf c}}
\def\qbar{\bar{\bf q}}

\input epsf.sty
\newdimen\psfigsize
\def\psfigure#1 #2 #3 #4 #5{
    \begin{figure}[tbh]
      \begin{center}
      \vbox{
        \null\vskip-0.2in\hskip#2
        \epsfxsize=#1
        \epsfbox{#4}
        \vskip -0.3in
        \caption {#5 \label{#3}}
        \vskip 0.0 true in plus 0.3 true in
      }
      \end{center}
   \end{figure}
}
\usepackage{graphicx}
\begin{document}
\thispagestyle{empty}
\begin{flushright}
NTUTH-06-505A \\
March 2006 \\
\end{flushright}
\vskip 2.5truecm
\centerline{{\LARGE $ X(3872) $ in lattice QCD with
exact chiral symmetry}}
\vskip 1.0truecm
\centerline{{\bf Ting-Wai~Chiu$^{1}$, Tung-Han~Hsieh$^{2}$}}
\vskip2.0ex
\centerline{$^1\hskip-3pt$ \it
Department of Physics, National Taiwan University,}
\vskip1.0ex
\centerline{\it Taipei, 10617, Taiwan}
\vskip2.0ex
\centerline{$^2\hskip-3pt$ \it
Physics Section, Commission of General Education,}
\vskip1.0ex
\centerline{\it National United University, Miao-Li, 36003, Taiwan}
\vskip1.0ex
\vskip2.0ex
\centerline{\bf (TWQCD Collaboration)}
\vskip 1cm
\bigskip \nopagebreak \begin{abstract}

\noindent

We investigate the mass spectrum of $ 1^{++} $ exotic 
mesons with quark content $ (\c\q\cbar\qbar) $,  
using molecular and diquark-antidiquark operators,  
in quenched lattice QCD with exact chiral symmetry.
For the molecular operator 
$ \{ (\qbar\gamma_i\c)(\cbar\gamma_5\q)-
     (\cbar\gamma_i\q)(\qbar\gamma_5\c) \} $ 
and the diquark-antidiquark operator 
$ \{ (\q^T C \gi \c )(\qbar C \gamma_5 \cbar^T) 
    -(\qbar C \gi^T \cbar^T)(\q^T C \gamma_5 \c) \} $,  
both detect a resonance with mass around $ 3890 \pm 30 $ MeV
in the limit $ m_q \to m_u $,   
which is naturally identified with $ X(3872) $. 
Further, heavier exotic meson resonance with $ J^{PC} = 1^{++} $ 
is also detected, with quark content $ (\c\s\cbar\sbar) $ around 
$ 4100 \pm 50 $ MeV.


Charm Mesons

\end{abstract}
\vskip 1.5cm 
\newpage\setcounter{page}1

\section{Introduction}

Since the discovery of $ D_s(2317) $ by BaBar in April 2003, 
a series of new heavy mesons\footnote{ 
For recent reviews of these new heavy mesons,  
see, for example, 
Refs. \cite{Swanson:2006st,Barnes:2005zy,Quigg:2005tv,
Rosner:2005gf}, and references therein.} 
with open-charm and closed-charm have been observed by 
Belle, CDF, CLEO, BaBar, and BES. 
Among these new heavy mesons, the narrow charmonium-like state $ X(3872) $ 
(with width $ < 2.3 $ MeV) first observed by Belle \cite{Choi:2003ue}
in the exclusive decay $ B^{\pm} \to K^\pm X \to K^\pm \pi^+ \pi^- J/\psi $ 
seems to be the most remarkable.  
The evidence of $ X(3872) $ has been confirmed by three experiments in 
three decay and two production channels 
\cite{Acosta:2003zx,Abazov:2004kp,Aubert:2004ns}.
Recently, the decay $ X(3872) \to \gamma J/\psi $ has been observed by 
Belle \cite{Abe:2005ix}, 
which implies that the charge conjugation of $ X(3872) $ is positive. 
Further, $ X(3872) $ is unlikely a vector $ 1^{--} $ state, 
since the bound $ \Gamma(e^+ e^-) Br(X \to \pi^+\pi^- J/\Psi) < 10 eV $ 
at $ 90\% $ C.L. has been obtained \cite{Yuan:2003yz},  
with the data collected by BES at $ \sqrt{s} = 4.03 $ GeV.  
Now, the quantum numbers $ 0^{++} $ and $ 0^{-+} $ are ruled out  
based on the angular correlations in the $ \pi^+\pi^- J/\psi $ system   
\cite{Abe:2005iy},  
also $ 2^{-+} $ and $ 1^{-+} $ are strongly disfavored according to the 
dipion mass distribution \cite{Abe:2005iy}.   
Thus it is likely that $ X(3872) $ possesses 
$ J^{PC} = 1^{++} $ \cite{Abe:2005iy}. 

Theoretically, one can hardly interpret $ X(3872) $ as $ 2P $ or $ 1D $ 
states in the charmonium spectrum, in view of its extremely narrow width. 
Thus it is most likely an exotic (non-$q\bar q $) meson 
(e.g., molecule, diquark-antidiquark, and hybrid meson).  
Now the central question is whether  
the spectrum of QCD possesses a resonance around 3872 MeV 
with $ J^{PC} = 1^{++} $.  

In this paper, we investigate the mass spectra of molecular
and diquark-antidiquark interpolating operators 
whose lowest-lying states having $ J^{PC} = 1^{++} $,  
in lattice QCD with exact chiral symmetry
\cite{Kaplan:1992bt,Narayanan:1995gw,Neuberger:1997fp,
      Ginsparg:1981bj,Chiu:2002ir}. 
This study follows our recent investigation on the 
mass spectrum of exotic mesons with $ J^{PC} = 1^{--} $ 
\cite{Chiu:2005ey}, which suggests that $ Y(4260) $ \cite{Aubert:2005rm}
is in the spectrum of QCD, with quark content $ (\c\u\cbar\ubar) $.  

For two lattice volumes $ 24^3 \times 48 $ and $ 20^3 \times 40 $, 
each of 100 gauge configurations generated with single-plaquette action 
at $ \beta = 6.1 $, we compute point-to-point quark propagators   
for 30 quark masses in the range $ 0.03 \le m_q a \le 0.80 $, 
and measure the time-correlation functions of the exotic meson operators 
which can overlap with $ X(3872) $.  
The inverse lattice spacing $ a^{-1} $
is determined with the experimental input of $ f_\pi $,
while the strange quark bare mass $ m_s a = 0.08 $, and the charm
quark bare mass $ m_c a = 0.80 $ are fixed such that the masses of
the corresponding vector mesons are in good agreement with  
$ \phi(1020) $ and $ J/\psi(3097) $ respectively 
\cite{Chiu:2005zc}. 

Note that we are working in the quenched approximation which 
in principle is unphysical. However, our previous results on 
charmed baryon masses \cite{Chiu:2005zc}, 
and also charmed meson masses and decay constants (theoretical predictions) 
\cite{Chiu:2005ue} turn out to be in good 
agreement with the experimental values. This suggests that it is
plausible to use the quenched lattice QCD with exact chiral symmetry
to investigate the mass spectra of the exotic meson operators 
constructed in this paper, as first steps toward the unquenched calculations.
The systematic error due to quenching can be determined
only after we can repeat the same calculation with unquenched
gauge configurations. However, the Monte Carlo simulation
of unquenched gauge configurations for lattice QCD with exact
chiral symmetry, on the lattices $ 20^3 \times 40 $ and $ 24^3 \times 48 $
at $ \beta = 6.1 $, still remains a challenge to the lattice community. 
Thus, in this paper, we proceed with the quenched approximation,
assuming that the quenching error does not change our conclusions
dramatically, in view of the good agreement between our previous
quenched mass spectra of charmed hadrons \cite{Chiu:2005zc,Chiu:2005ue}
and the experimental values.

\section{The Molecular Operators}

In this section, we construct four molecular operators with quark 
content ($ \c\q\cbar\qbar $) such that the lowest-lying state of 
each operator has $ J^{PC} = 1^{++} $. Then we compute the 
time correlation function of each operator, and extract the
mass of its lowest-lying state. Explicitly, these molcular operators are: 
\bea
\label{eq:DVDA}
M_1 \EQ \frac{1}{\sqrt{2}} \left\{ (\qbar\gi\c)(\cbar\gamma_5\q)
                                  -(\cbar\gi\q)(\qbar\gamma_5\c) \right\} \\
\label{eq:etacIrho5}
M_2 \EQ (\qbar\gamma_5\gi\q)(\cbar\c) \\         
\label{eq:D5VDIS}
M_3 \EQ \frac{1}{\sqrt{2}} \left\{ (\qbar\gamma_5\gamma_i\c)(\cbar\q)
              +(\cbar\gamma_5\gamma_i\q)(\qbar\c) \right\}  \\
\label{eq:J5pi}
M_4 \EQ (\cbar\gamma_5\gi\c)(\qbar\q)          
\eea

\begin{figure}[htb]
\begin{center}
\includegraphics*[height=10cm,width=8cm]{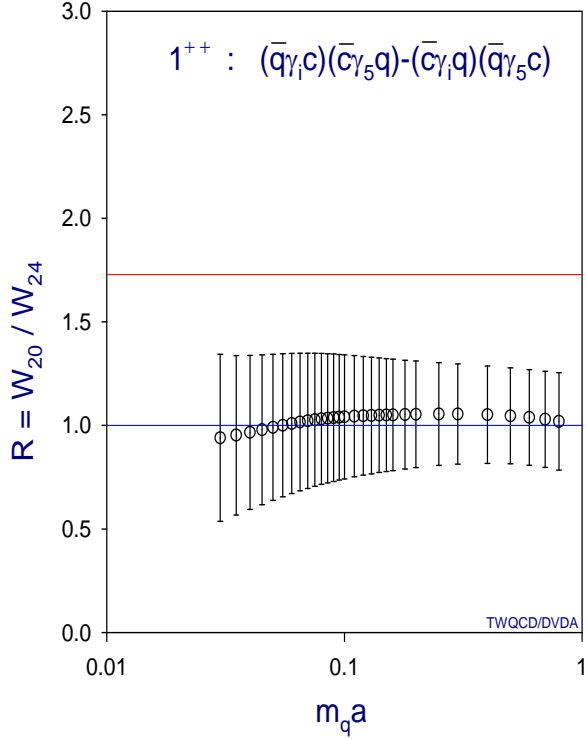}
\caption{
The ratio of spectral weights of the lowest-lying state
of the molecular operator $ M_1 $,
for $ 20^3 \times 40 $ and $ 24^3 \times 48 $ lattices at $ \beta = 6.1 $.
The upper-horizontal line $ R = (24/20)^3 = 1.728 $,
is the signature of 2-particle scattering state,
while the lower-horizontal line $ R = 1.0 $ is the signature
of a resonance.}
\label{fig:sw2024_DVDA}
\end{center}
\end{figure}

\begin{figure}[htb]
\begin{center}
\includegraphics*[height=10cm,width=8cm]{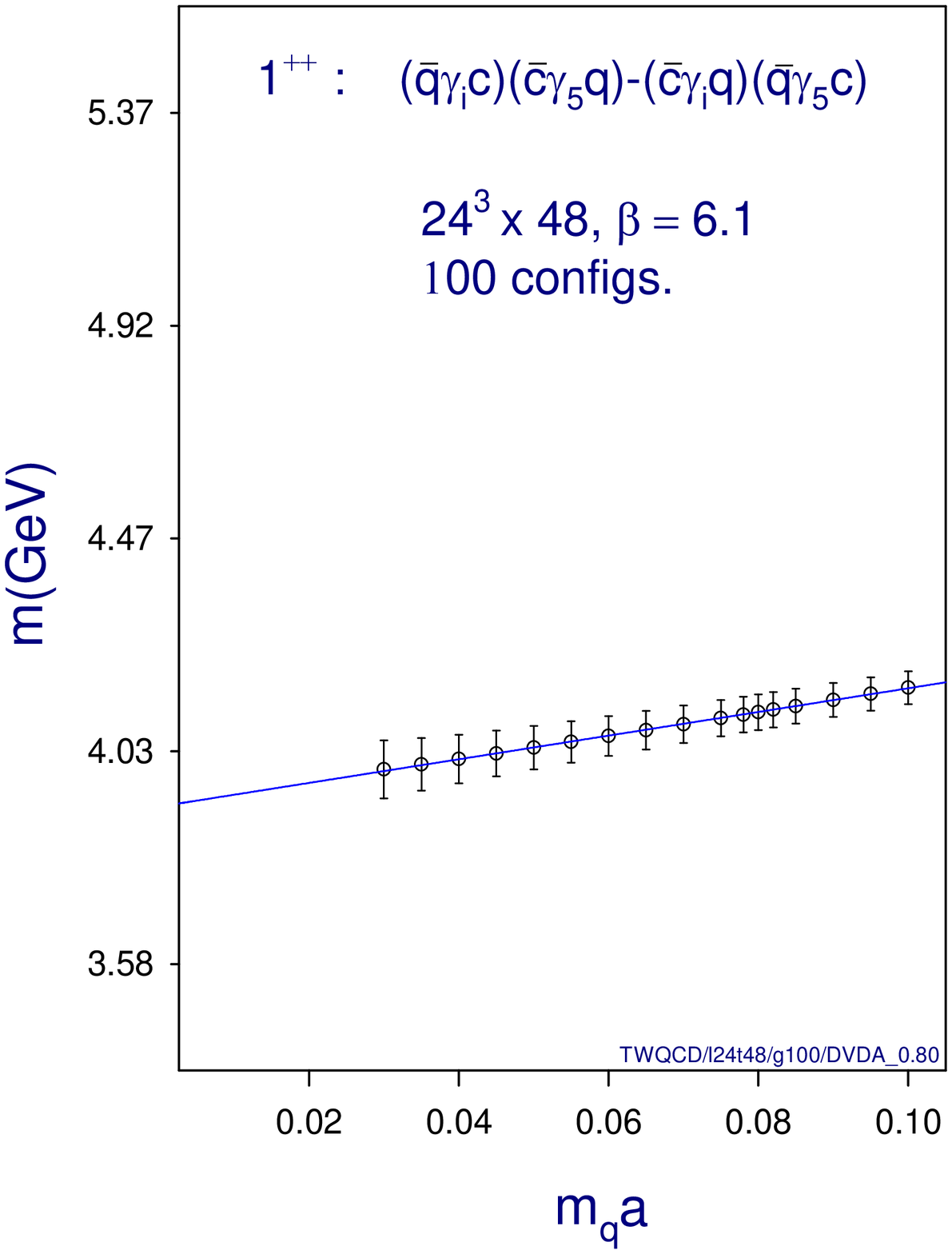}
\caption{
The mass of the lowest-lying state
of $ M_1 $ versus the quark mass $ m_q a $, 
on the $ 24^3 \times 48 $ lattice at $ \beta = 6.1 $.
The solid line is the linear fit.}
\label{fig:mass_DVDA}
\end{center}
\end{figure}

\begin{figure}[htb]
\begin{center}
\begin{tabular}{@{}cc@{}}
\includegraphics*[height=9cm,width=7cm]{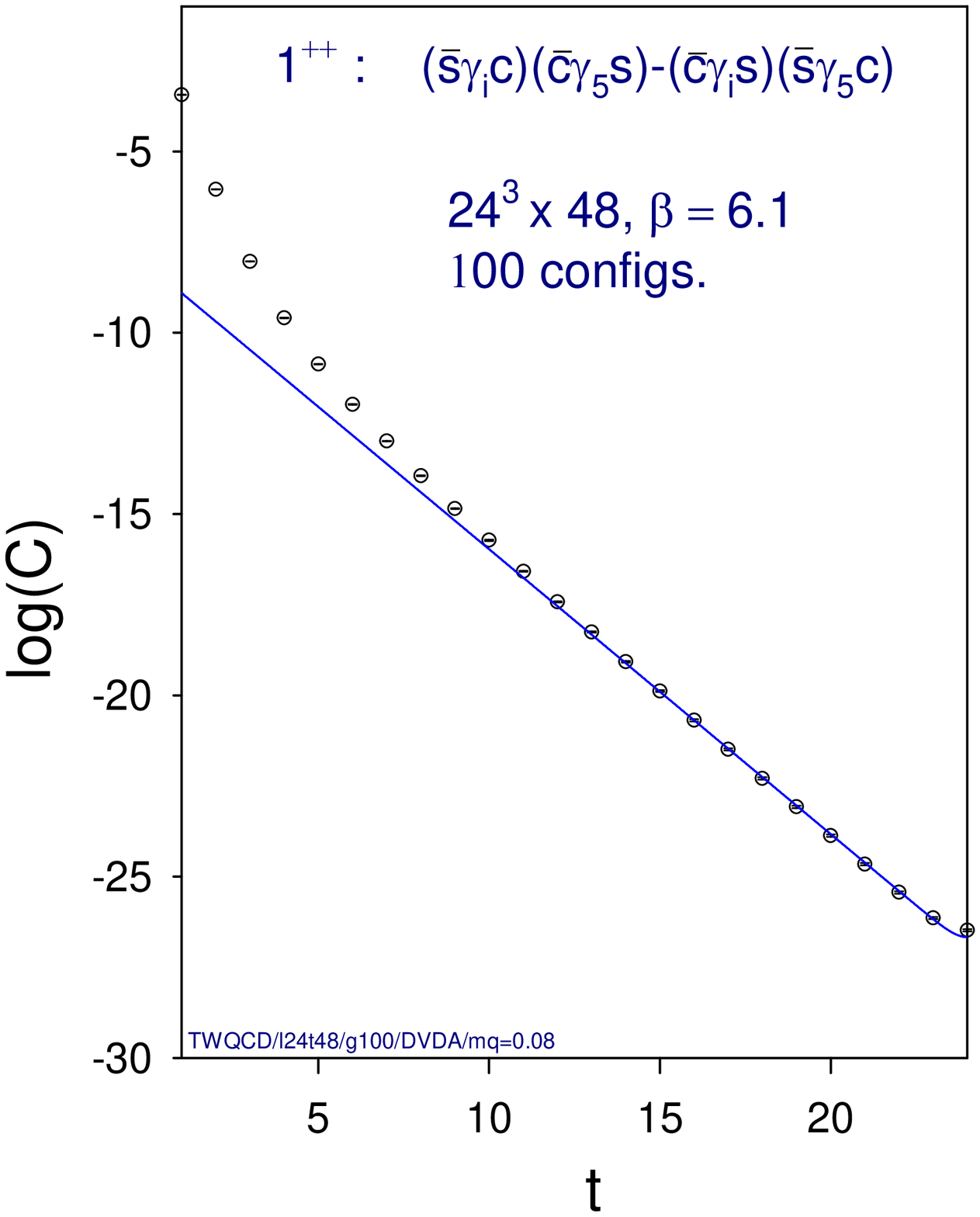}
&
\includegraphics*[height=9cm,width=7cm]{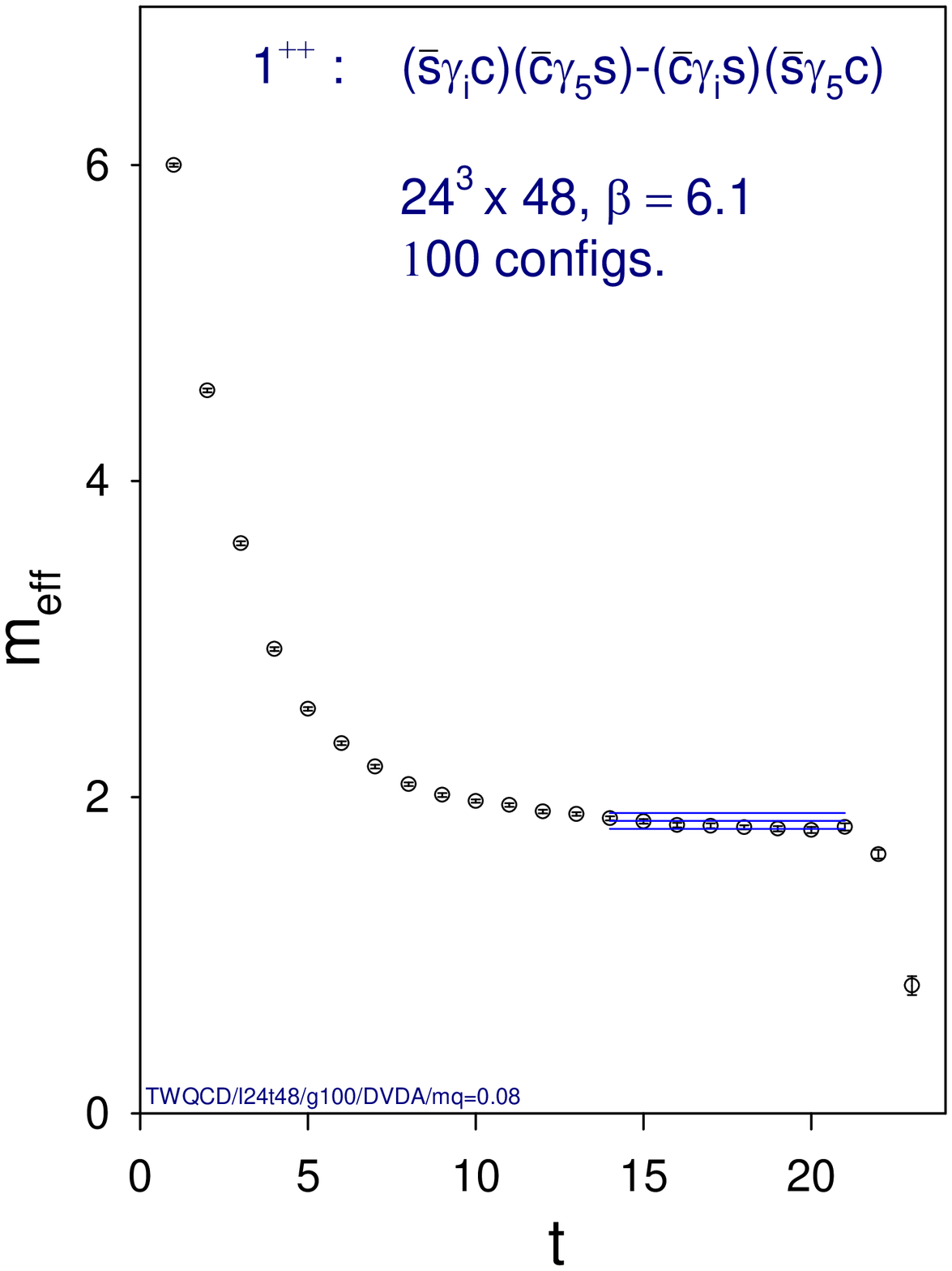}
\\ (a) & (b)
\end{tabular}
\caption{
(a) The time-correlation function $ C(t) $ of the lowest-lying state
of $ M_1 $ for $ m_q = m_s = 0.08 a^{-1} $,  
on the $ 24^3 \times 48 $ lattice at $ \beta = 6.1 $.
The solid line is the hyperbolic-cosine fit for $ t \in [14,21] $.
(b) The effective mass $ M_{eff}(t) = \ln [C(t)/C(t+1)]  $
of $ C(t) $ in Fig.\ \ref{fig:DVDA_008}a.
}
\label{fig:DVDA_008}
\end{center}
\end{figure}

The time-correlation function\footnote{
Here we have neglected the 
$ \c\cbar $ and $ \q\qbar $ annihilation diagrams 
such that $ C(t) $ does not overlap with any conventional meson  
($ \c\cbar $ or $ \q\qbar $) states. Also,   
$ C(t) $ has been averaged over $ C_i $ (with $ \gamma_i $) 
for $ i=1,2,3 $, where in each case,
the ``forward-propagator" $ C_i(t) $ and ``backward-propagator"
$ C_i(T-t) $ are averaged to increase the statistics.
The same strategy is applied to all time-correlation functions
in this paper.} 
\BAN
C_{M}(t)=\sum_{\vec{x}} \left< M(\vec{x},t) M^\dagger(\vec{0},0) \right>
\EAN
is measured for each gauge configuration, and its average over all 
gauge configurations is fitted to the usual formula 
\BAN
\frac{Z}{2 m a } [ e^{-m a t} + e^{-m a (T-t)} ]
\EAN
to extract the mass $ m a $ of the lowest-lying state 
and its spectral weight
\BAN
W = \frac{Z}{2 m a } \ .
\EAN
Theoretically, if this state is a genuine resonance, then its mass $ m a $  
and spectral weight $ W $ should be almost constant for 
any lattices with the same lattice spacing. On the 
other hand, if it is a 2-particle scattering state, then its mass 
$ m a $ is sensitive to the lattice volume, and its spectral
weight is inversely proportional to the spatial volume for lattices
with the same lattice spacing.
In the following, we shall use the ratio of the spectral weights on 
two spatial volumes $ 20^3 $ and $ 24^3 $ with the same lattice spacing 
($\beta = 6.1 $) to discriminate whether any hadronic state under 
investigation is a resonance or not.

In Fig. \ref{fig:sw2024_DVDA}, the ratio ($ R=W_{20}/W_{24} $) 
of spectral weights of the lowest-lying state extracted from 
the time-correlation function of $ M_1 $ on the 
$ 20^3 \times 40 $ and $ 24^3 \times 48 $ lattices is plotted 
versus the quark mass $ m_q a \in [0.03, 0.80] $.   
(Here the quark fields $ \q $ and $ \qbar $ are always 
taken to be different from $ \c $ and $ \cbar $, 
even in the limit $ m_q \to m_c $.)
Evidently, $ R \simeq 1.0 $ for the entire range of quark masses,   
which implies that there exist $ J^{PC} = 1^{++} $ resonances, 
with quark content $ (\c\s\cbar\sbar) $, 
and $ (\c\u\cbar\ubar) $ respectively.

In Fig. \ref{fig:mass_DVDA}, the mass of the lowest-lying state 
extracted from the molecular operator $ M_1 $ is plotted versus $ m_q a $.  
In the limit $ m_q \to m_{u} \simeq 0.00265 a^{-1} $ 
(corresponding to $ m_\pi = 135 $ MeV), it gives $ m = 3895(27) $ MeV, 
which is in good agreement with the mass of $ X(3872) $.

For $ m_q = m_s = 0.08 a^{-1} $, the time-correlation function 
and effective mass of $ M_1 $ are plotted in Fig. \ref{fig:DVDA_008}a  
and Fig. \ref{fig:DVDA_008}b respectively.
With single exponential fit, it gives 
$m[(\sbar\gi\c)(\cbar\gamma_5\s)-(\cbar\gi\s)(\sbar\gamma_5\c)]=4109(21)$ MeV.

Next we turn to the molecular operators $ M_2 $, $ M_3 $, and $ M_4 $.
In each of these cases, the ratio of spectral weights ($ R = W_{20}/W_{24} $) 
behaves as $ R \simeq 1.0 $ for $ m_q a > 0.05 $, but 
deviates from 1.0 with large errors as $ m_q \to m_{u} $.  
This seems to suggest that $ M_2 $, $ M_3 $ and $ M_4 $ have little overlap 
with the resonance detected by $ M_1 $ as $ m_q \to m_u $. 
We suspect that this is 
due to quenching artifacts as $ m_q \to m_{u} $,
mostly coming from the scalar meson $ (\qbar \q) $, 
or $ (\cbar \q) $, or $ (\qbar \c) $, as well as the pseudovector 
$ (\qbar \gamma_5 \gamma_i \q) $. 
However, all molecular operators give compatible 
masses as $ m_q \to m_u $. This is also the case for
the resonance at $ m_q = m_s $.

\section{The Diquark-Antidiquark Operator}

\begin{figure}[htb]
\begin{center}
\includegraphics*[height=10cm,width=8cm]{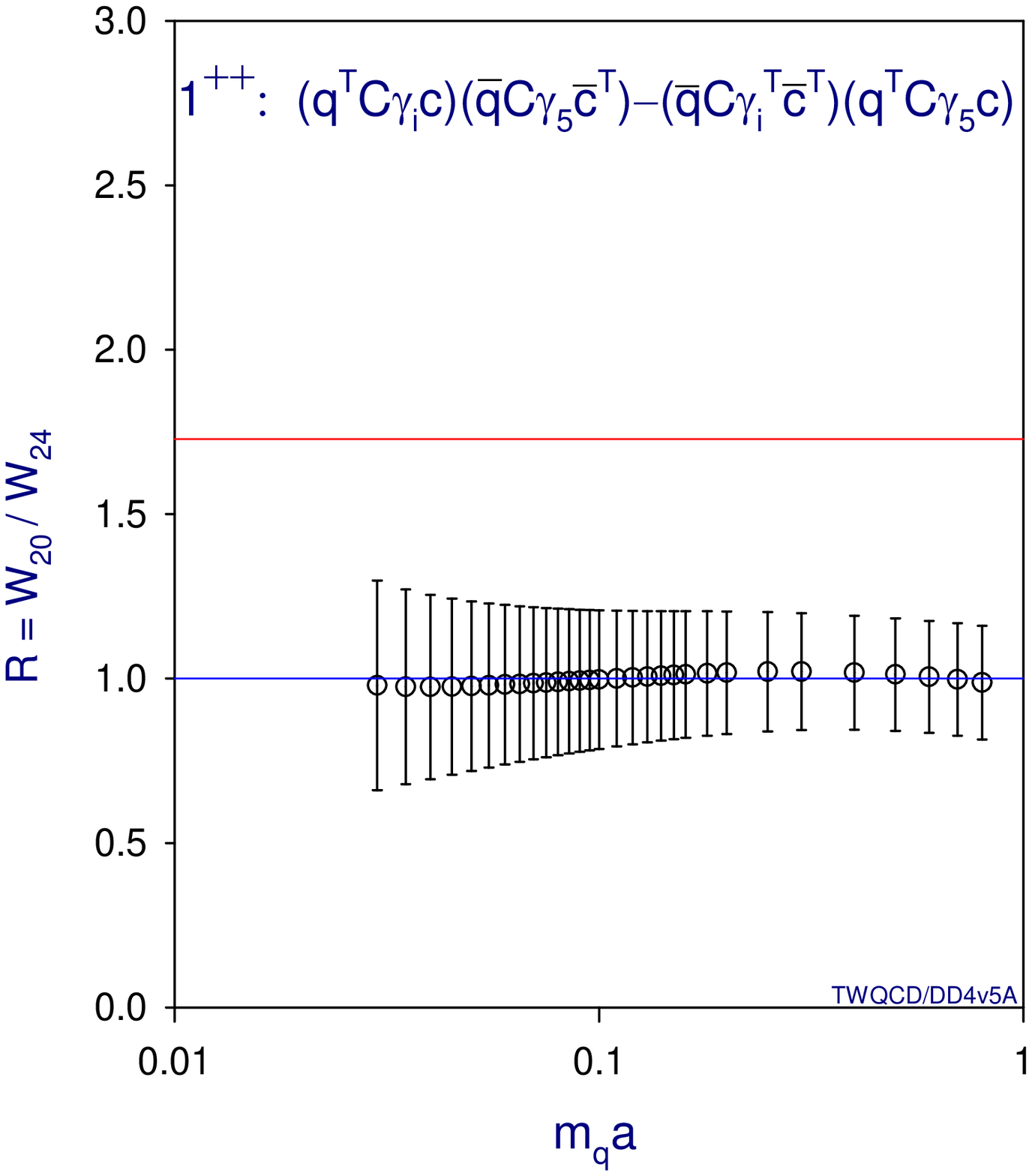}
\caption{
The ratio of spectral weights of the lowest-lying state
of diquark-antidiquark operator $ X_4 $, 
for $ 20^3 \times 40 $ and $ 24^3 \times 48 $ lattices at $ \beta = 6.1 $.
The upper-horizontal line $ R = (24/20)^3 = 1.728 $,
is the signature of 2-particle scattering state,
while the lower-horizontal line $ R = 1.0 $ is the signature
of a resonance.}
\label{fig:sw2024_DD4V5A}
\end{center}
\end{figure}

\begin{figure}[htb]
\begin{center}
\includegraphics*[height=10cm,width=8cm]{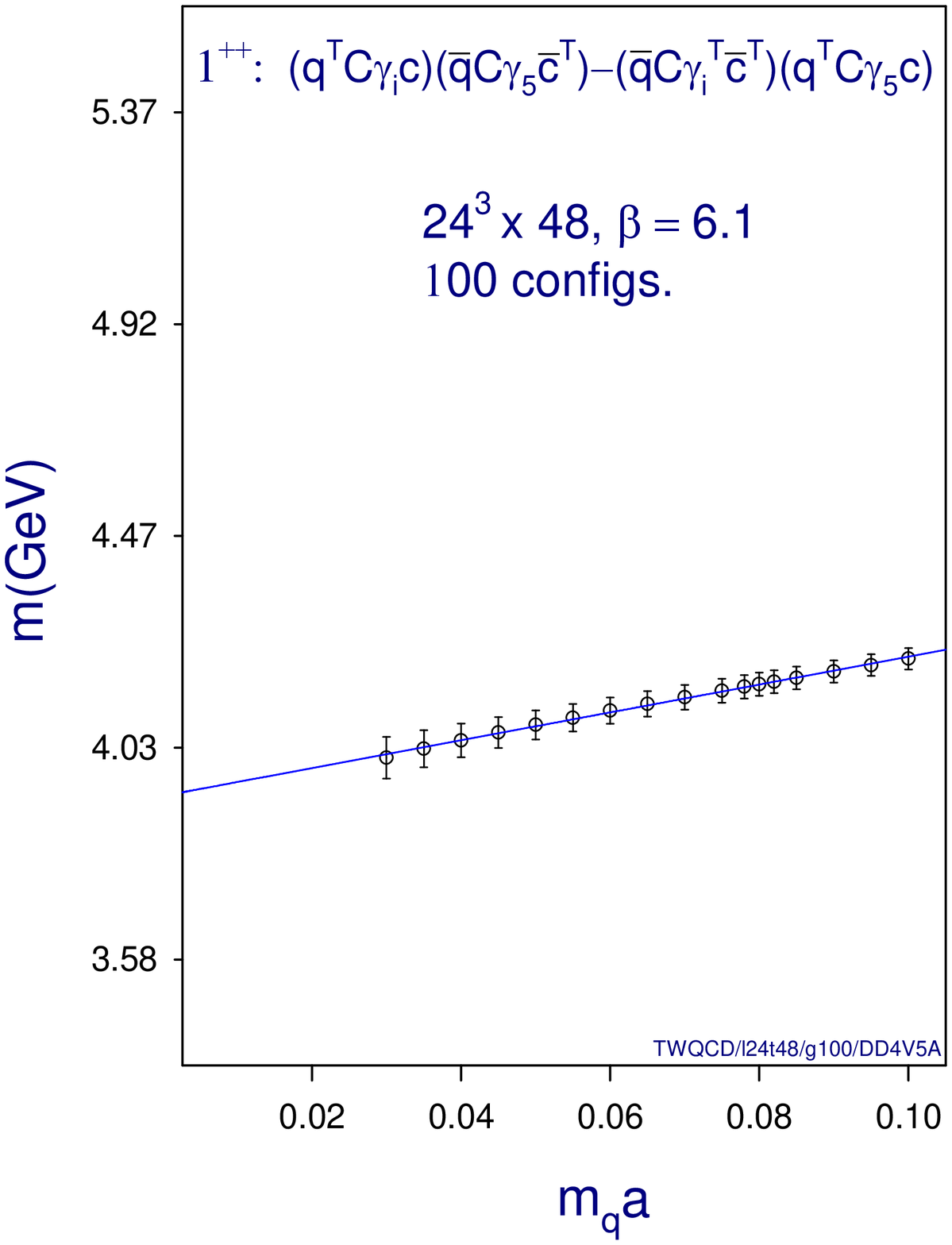}
\caption{
The mass of the lowest-lying state of the diquark-antidiquark operator
$ X_4 $ versus the quark mass $ m_q a $,
on the $ 24^3 \times 48 $ lattice at $ \beta = 6.1 $.
The solid line is the linear fit.}
\label{fig:mass_DD4V5A}
\end{center}
\end{figure}

\begin{figure}[htb]
\begin{center}
\begin{tabular}{@{}cc@{}}
\includegraphics*[height=9cm,width=7cm]{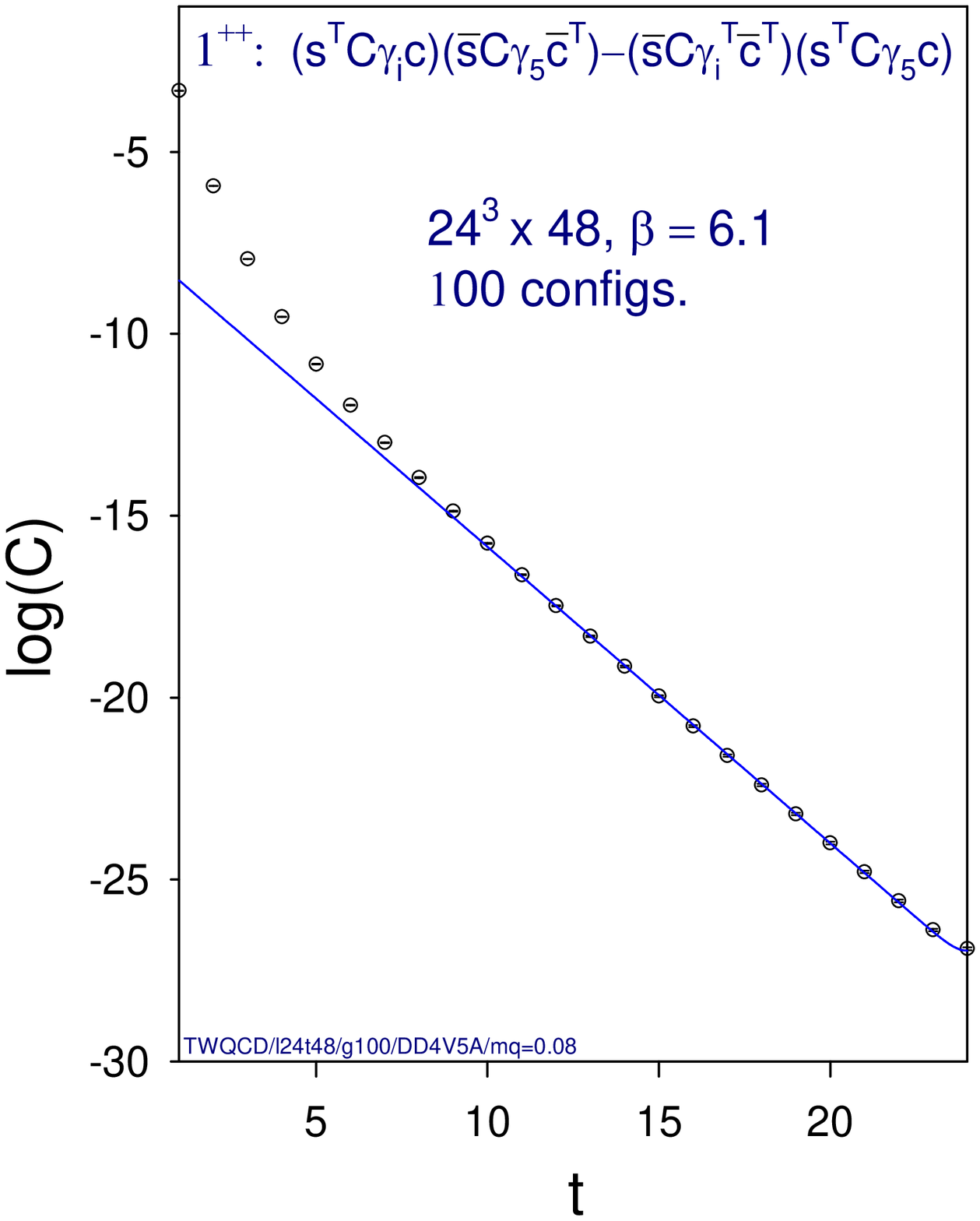}
&
\includegraphics*[height=9cm,width=7cm]{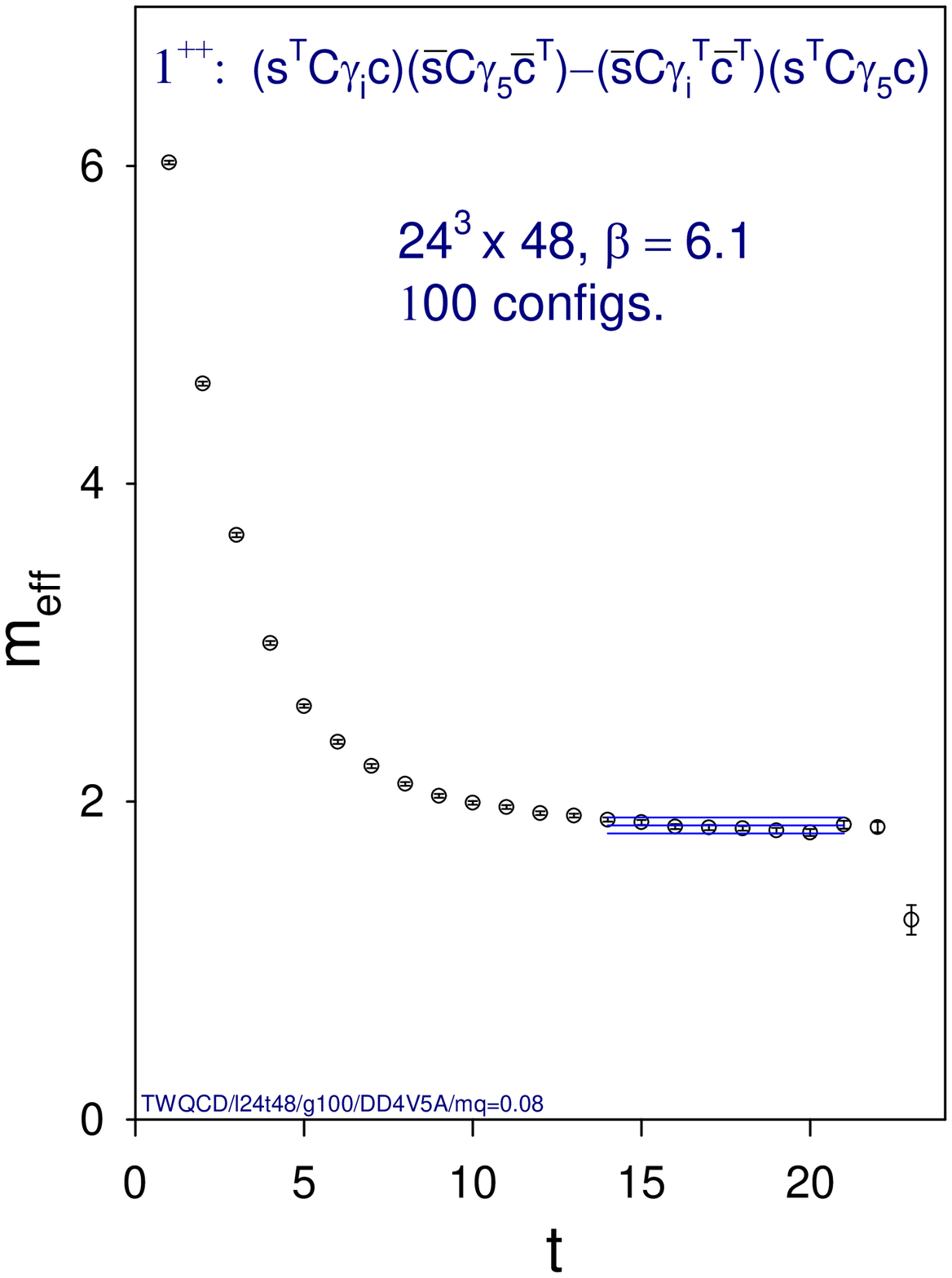}
\\ (a) & (b)
\end{tabular}
\caption{
(a) The time-correlation function $ C(t) $ of the lowest-lying state
of $ X_4 $ for $ m_q = m_s = 0.08 a^{-1} $,
on the $ 24^3 \times 48 $ lattice at $ \beta = 6.1 $.
The solid line is the hyperbolic-cosine fit for $ t \in [14,21] $.
(b) The effective mass $ M_{eff}(t) = \ln [C(t)/C(t+1)]  $
of $ C(t) $ in Fig.\ \ref{fig:DD4V5A_008}a.
}
\label{fig:DD4V5A_008}
\end{center}
\end{figure}


We construct the diquark-antidiquark operator with $ J^{PC} = 1^{++} $ as 
\bea
\label{eq:DD4V5A}
X_4(x)
\EQ \frac{1}{\sqrt{2}} \left\{ 
   (\q^T C \gi \c )_{xa}(\qbar C \gamma_5 \cbar^T)_{xa} 
  -(\qbar C \gi^T \cbar^T)_{xa} (\q^T C \gamma_5 \c)_{xa} \right\} 
\eea
where $ C $ is the charge conjugation operator satisfying
$ C \gamma_\mu C^{-1} = -\gamma_\mu^T $ and
$ (C \gamma_5)^T=-C\gamma_5 $. Here the 
``diquark" operator $ (\q^T \Gamma \Q)_{xa} $ for any Dirac matrix 
$ \Gamma $ is defined as
\bea
\label{eq:diquark}
({\q}^T \Gamma {\Q} )_{xa} \equiv \epsilon_{abc} 
 {\q}_{x\alpha b} \Gamma_{\alpha\beta} {\Q}_{x\beta c}
\eea
where 
$ x $, $ \{ a,b,c \} $ and $ \{ \alpha, \beta \} $
denote the lattice site, color, and Dirac indices respectively,  
and $ \epsilon_{abc} $ is the completely antisymmetric tensor. 
Thus the diquark (\ref{eq:diquark}) transforms like color anti-triplet.  
For $ \Gamma = C \gamma_5 $, it transforms like $ J^P = 0^{+} $,  
while for $ \Gamma = C \gamma_i $ ($i=1,2,3 $), it 
transforms like $ 1^{+} $. 

In Fig. \ref{fig:sw2024_DD4V5A}, the ratio ($ R=W_{20}/W_{24} $)
of spectral weights of the lowest-lying state extracted from
the time-correlation function of $ X_4 $ on the $ 20^3 \times 40 $
and $ 24^3 \times 48 $ lattices is plotted
versus the quark mass $ m_q a \in [0.03, 0.8] $.
Evidently, $ R \simeq 1.0 $ for the entire range of quark masses,   
which implies that there exist $ J^{PC} = 1^{++} $ resonances, 
with quark content 
$ (\c\s\cbar\sbar) $, 
and $ (\c\u\cbar\ubar) $ respectively.

In Fig. \ref{fig:mass_DD4V5A}, the mass of the lowest-lying state
of the diquark-antidiquark operator $ X_4 $
is plotted versus $ m_q a $. In the limit $ m_q \to m_{u} $, it gives
$ m = 3891(17) $ MeV, which is in good agreement
with the mass of $ X(3872) $.

For $ m_q = m_s = 0.08 a^{-1} $, 
the time-correlation function and effective mass of the 
diquark-antidiquark operator are plotted in Fig. \ref{fig:DD4V5A_008}.  
The mass of the lowest-lying state is 4134(19) MeV.

\begin{table}
\begin{center}
\begin{tabular}{c|c|c}
Operator & Mass (MeV) & R/S  \\
\hline
\hline
$\frac{1}{\sqrt{2}}[ (\ubar \gamma_i \c)(\cbar \gamma_5 \u)
                    -(\cbar \gamma_i \u)(\ubar \gamma_5 \c) ] $
        &   3895(27)(35) &   R  \\
$ \frac{1}{\sqrt{2}}[ (\sbar \gamma_i \c)(\cbar \gamma_5 \s)
                     -(\cbar \gamma_i \s)(\sbar \gamma_5 \c) ]  $
        &   4109(21)(32) &    R       \\
\hline
$ \frac{1}{\sqrt{2}}\left\{(\u^T C\gamma_i \c)(\ubar C\gamma_5\cbar^T)
  -(\u^T C \gamma_5 \c)(\ubar C\gamma_i^T \cbar^T) \right\} $      &
 3891(17)(21)  & R \\
$ \frac{1}{\sqrt{2}} \left\{ (\s^T C\gamma_i\c)(\sbar C\gamma_5\cbar^T)
  -(\s^T C \gamma_5 \c)(\sbar C\gamma_i^T \cbar^T) \right\} $      &
 4134(19)(25) & R \\
\hline
\hline
\end{tabular}
\caption{Mass spectra of the molecular operator $ M_1 $
and the diquark-antidiquark operator $ X_4 $ with $ J^{PC} = 1^{++} $.
The last column R/S denotes resonance (R) or scattering (S) state.}
\label{tab:mass_summary}
\end{center}
\end{table}

\section{Summary and Discussions}

In this paper, we have investigated the mass spectra of 
several interpolating operators (i.e.,  
the molecular operators $ M_1 $, $ M_2 $, $ M_3 $, and $ M_4 $, 
and the diquark-antidiquark operator $ X_4 $) with the lowest-lying 
$ J^{PC} = 1^{++} $, in quenched lattice QCD with exact chiral symmetry. 
Our results for $ M_1 $ and $ X_4 $ are summarized in 
Table \ref{tab:mass_summary}, 
where in each case, the first error is statistical, and 
the second one is our estimate of combined systematic uncertainty 
including those coming from:
(i) possible plateaus (fit ranges) with $ \chi^2/d.o.f. < 1 $;
(ii) the uncertainties in the strange quark mass and the charm quark mass;
(iii) chiral extrapolation (for the entries containing u/d quarks); and
(iv) finite size effects (by comparing results of two lattice sizes).
Note that we cannot estimate the discretization error
since we have been working with one lattice spacing.
Even though lattice QCD with exact chiral symmetry does not have
$ O(a) $ and $ O(ma) $ lattice artifacts, the $ O(m^2 a^2) $ effect
might turn out to be not negligible for $ m_c a = 0.8 $.

Evidently, both the molecular operator $ M_1 $
and the diquark-antidiquark operator $ X_4 $ 
detect a $ 1^{++} $ resonance around $ 3890 \pm 30 $ MeV  
in the limit $ m_q \to m_u $, which is naturally identified 
with $ X(3872) $. 
This suggests that $ X(3872) $ is indeed in the spectrum of QCD, 
with quark content ($\c\u\cbar\ubar$), and $ J^{PC} = 1^{++} $. 

Now, in the quenched approximation, our results  
suggest that $ X(3872) $ has good overlap with the molecular 
operator $ M_1 $ as well as the diquark-antidiquark operator $ X_4 $.
This is in contrast to the case of $ Y(4260) $ in our recent 
study \cite{Chiu:2005ey}, in which $ Y(4260) $ seems to have better 
overlap with the molecular operator 
$ \{ (\qbar\gamma_5\gamma_i\c)(\cbar\gamma_5\q)-
     (\cbar\gamma_5\gamma_i\q)(\qbar\gamma_5\c) \} $
than any diquark-antidiquark operators.
This seems to suggest that $ X(3872) $ is more tightly bound than
$ Y(4260) $. It would be interesting to see whether this 
picture persists even for unquenched QCD.

Finally, for $ m_q = m_s $, 
heavier exotic meson resonance with $ J^{PC} = 1^{++} $ is also detected,   
with quark content ($\c\s\cbar\sbar$) around $ 4100 \pm 50 $ MeV, 
which serves as a prediction from lattice QCD with exact chiral symmetry.

\bigskip
\bigskip

\flushpar
{\bf Acknowledgement}

\noindent

\bigskip

This work was supported in part by the National Science Council,
Republic of China, under the Grant No. NSC94-2112-M002-016 (T.W.C.),  
and Grant No. NSC94-2119-M239-001 (T.H.H.), and by  
the National Center for High Performance Computation at Hsinchu, 
and the Computer Center at National Taiwan University.

\bigskip
\bigskip

\vfill\eject

\end{document}